\input harvmac
\input epsf
\batchmode
\font\bbbfont=msbm10
\errorstopmode
\newif\ifamsf\amsftrue
\ifx\bbbfont\nullfont
  \amsffalse
\fi
\ifamsf
\def\IR{\hbox{\bbbfont R}}
\def\IZ{\hbox{\bbbfont Z}}
\def\IF{\hbox{\bbbfont F}}
\def\IP{\hbox{\bbbfont P}}
\else
\def\IR{\relax{\rm I\kern-.18em R}}
\def\IZ{\relax\ifmmode\hbox{Z\kern-.4em Z}\else{Z\kern-.4em Z}\fi}
\def\IF{\relax{\rm I\kern-.18em F}}
\def\IP{\relax{\rm I\kern-.18em P}}
\fi
\def\IC{{\bf C}}
\def\np#1#2#3{Nucl. Phys. B {#1} (#2) #3}

\def\plb#1#2#3{Phys. Lett. B {#1} (#2) #3}
\def\prl#1#2#3{Phys. Rev. Lett. {#1} (#2) #3}
\def\physrev#1#2#3{Phys. Rev. D {#1} (#2) #3}

\def\ev#1{\langle#1\rangle}

\def\M{{\cal M}}

\def\f#1#2{\textstyle{#1\over #2}}

\def\N20{${\cal N}=(2,0)$}

\nref\GW{J. Goldstone and F. Wilczek, \prl{47}{1981}{986}.}
\nref\EWWZ{E. Witten, \np{223}{1983}{422}; \np{223}{1983}{433}.}
\nref\EDEF{E. D 'Hoker and E. Farhi, \np{248}{1984}{59}.}
\nref\manohar{A.V. Manohar, hep-th/9805144, \prl{81}{1998}{1558}.}
\nref\Rub{V.A. Rubakov, hep-th/9812128; S.L. Dubovsky, D.S. Gorbunov,
M.V. Libanov, and V.A. Rubakov, hep-th/9903155.}
\lref\thooft{G. t'Hooft, {\it Recent Developments in Gauge Theories},
eds. G. 't Hooft et. al., Plenum Press, NY, 1980.}
\lref\malda{J.M. Maldacena, hep-th/9711200, Adv. Theor. Math. Phys. 2
(1998) 231.}
\lref\nahm{W. Nahm, \np{135}{1978}{149}.}
\lref\minwalla{S. Minwalla, hep-th/9712074, Adv. Theor. Math. Phys. 2
(1998) 781.}
\lref\wcomments{E. Witten, hep-th/9507121, in proc. of Strings 95,
Eds. Bars et. al.}
\lref\malda{J. M. Maldacena, hep-th/9711200, Adv. Theor. Math. Phys.
2 (1998) 231.}
\lref\HSW{P.S. Howe, E. Sezgin, and P.C. West, hep-th/9702008, 
\plb{399}{1997}{49}.}
\lref\BHS{X. Bekaert, M. Henneaux, A. Sevrin, hep-th/9909094,
\plb{468}{1999}{228}.}
\lref\KT{I.R. Klebanov and A. Tseytlin, hep-th/9604089, \np{475}{1996}{164}.} 
\lref\HS{M. Henningson and K. Skenderis, hep-th/9806087, JHEP 9807
(1998) 023.}
\lref\HMM{J. Harvey, R. Minasian, and G. Moore, hep-th/9808060,
JHEP 9809 (1998) 004.}
\lref\FHMM{D. Freed, J. Harvey, R. Minasian, G. Moore, hep-th/9803205,
ATMP 2 (1998) 601.}
\lref\EWiv{E. Witten, hep-th/9610234, J. Geom. Phys. 22 (1997) 103.}
\lref\Hilton{{\it Encyclopaedia of Mathematics}, ed. M. Hazewinkel, 
Kluwer Academic Publishers, 1988.}
\lref\BC{R. Bott and A.S. Cattaneo, dg-ga/9710001, J. Diff. Geom,
48 (1998) 91.}
\lref\MQCD{E. Witten, hep-th/9706109, \np{507}{1997}{658}.}
\lref\AW{L. Alvarez-Gaume' and E. Witten, \np{234}{1983}{269}.}
\lref\Wiv{E. Witten, hep-th/9610234, J. Geom. Phys. 22 (1997) 103.}
\lref\ATKZ{A.A. Tseytlin and K. Zarembo, hep-th/9911246.}
\lref\Gubser{S.S. Gubser, hep-th/9807164, \physrev{59}{1999}{025006}.}
\lref\coleman{S. Coleman, Comm. Math. Phys. 31 (1973) 259.}
\lref\MSW{J.M. Maldacena, A. Strominger, and E. Witten,
hep-th/9711053, JHEP 12 (1997) 2.} 
\lref\OH{O. Aharony, hep-th/9604103, \np{476}{1996}{470}.}
\lref\OGLM{O. Ganor and L. Motl, hep-th/9803108, JHEP 9805 (1998) 009.}
\lref\HSW{P. Howe, E. Sezgin, P.C. West, hep-th/9702008,
\plb{399}{1997}{49}.}
\lref\DWZR{See e.g. M.B. Green, J.A. Harvey, 
and G. Moore, hep-th/9605033,
Class. Quant. Grav. 14 (1997) 47; Y.K.E. Cheung and Z. Yin, 
hep-th/9710206; and references cited therein.}
\lref\CBGT{C. Boulahouache and G. Thompson, hep-th/9801083, IJMP A13 
(1998) 5409. }
\lref\EDSW{E. D'Hoker and S. Weinberg, hep-ph/9409402, 
\physrev{50}{1994}{605};
E. D'Hoker, hep-th/9502162, \np{451}{1995}{725}.}
\lref\Getal{J. Gauntlett, C. Koehl, D. Mateos, P.K. Townsend, and M.
Zamaklar, hep-th/9903156, \physrev{60}{1999}{045004}.}
\Title{hep-th/0001205,  UCSD/PTH 00-02, IASSNS-HEP-00/06, RUNHETC-00-03}
{\vbox{\centerline{Anomaly Matching and a Hopf-Wess-Zumino}
\centerline{Term in 6d, \N20\ Field Theories
}}}
\medskip
\centerline{Kenneth Intriligator}
\vglue .5cm
\centerline{UCSD Physics Department}
\centerline{9500 Gilman Drive}
\centerline{La Jolla, CA 92093}
\vglue .25cm
\centerline{and}
\vglue .25cm
\centerline{Department of Physics\footnote{${}^*$}{visiting address, 
Winter, 2000.}}
\centerline{Rutgers University}
\centerline{Piscataway, NJ 08855-0849, USA}

\bigskip
\noindent

We point out that the low energy theory of 6d \N20\ field theories,
when away from the origin of the moduli space of vacua, necessarily
includes a new kind of Wess-Zumino term.  The form of this term is
related to the Hopf invariant associated with $\pi _7 (S^4)$.  The
coefficient of the Wess-Zumino term is fixed by an anomaly matching
relation for a global flavor symmetry.  For example, in the context of
a single M5 brane probe in the background of $N$ distant M5 branes,
the probe must have the Hopf-WZ term with coefficient proportional to
$N(N+1)$.  Various related checks and observations are made.  We also
point out that there are skyrmionic strings, and propose that they are
the $W$-boson strings.
\Date{1/00}

\newsec{Introduction}

Low energy effective field theories can have effects, generated by
integrating out massive fields, which do not decouple even when the
masses of these fields is taken to be infinite.  The classic example
is the Goldstone-Wilczek current \GW, which is generated by
integrating out fermions which get a mass via Yukawa coupling to a
scalar which gets an expectation value.  The GW current does not
depend on the masses of the integrated out fermions (as long as they
are non-zero), so it plays a role in the low energy theory even when
the masses are taken to be infinite (via large Yukawas or scalar
vevs).  A related effect is the Wess-Zumino term associated with
integrating out massive fields, which is often needed in the low
energy theory on symmetry grounds.  For example, a Wess-Zumino term,
with a particular coefficient, can be needed to reproduce the
contribution to the 't Hooft anomalies of integrated out fermions,
which get a mass via Yukawa couplings to a scalar which gets a vev.
The size of the WZ term can not depend on any parameters (e.g. Yukawa
or gauge couplings, vevs, etc.), or the RG scale: since its
coefficient must be quantized \EWWZ, it can not be renormalized.  See
e.g. \refs{\GW-\Rub} and references cited therein.

This paper will be concerned with flavor anomaly matching and
Wess-Zumino terms in the 6d \N20\ field theories. Much about these
theories, including how to properly formulate them as field theories,
remains mysterious.  They have the exotic property of having, rather
than ordinary gauge fields, interacting (somehow!, despite \BHS)
two-form gauge fields, with self-dual three-form field strengths.  The
existence of these field theories, as well as all of their known
properties, has come from string theory, where they occur in various
related contexts: the IR limit of the M5 or IIA NS-5 brane
world-volume theory, IIB string theory on a ALE singularity
\wcomments, M theory on $AdS_7\times S^4$ \malda, etc.

The 6d \N20\ theories are interesting, and worthy of further study,
both because of these connections to string theory and duality and, in
their own right, as field theories.  They are the maximally
supersymmetric conformal field theories, in the highest possible
dimension \refs{\nahm, \minwalla}, and other interesting theories can
be obtained by compactification and RG flow.  For example,
compactifying on a $T^2$ gives 4d ${\cal N}=4$ theories and makes
$SL(2,Z)$ electric-magnetic duality manifest, as the geometric symmetry
of the complex structure of $T^2$.  Instead compactifying on a $T^2$
with supersymmetry breaking boundary conditions leads to the theory
known as MQCD, which is hoped to be in the same universality class,
but more tractable than, ordinary, non-supersymmetric, pure glue, QCD 
\MQCD.

The 6d \N20\ theories are chiral, with an $SO(5)_R$ flavor symmetry.
Although the gauge fields are two-forms rather than one-forms, there
is a correspondence with non-Abelian groups $G$.  String theory
indicates that $G$ can be an arbitrary ADE group: $SU(N)$, $SO(2N)$,
or $E_{6,7,8}$.  (And $G=U(1)$ for the free \N20\ tensor multiplet.)
Upon compactification to lower dimensions, there is an ordinary gauge
symmetry with gauge group $G$.  In 6d, there is a moduli space of
supersymmetric vacua $\M=(\IR ^5)^{r(G)}/W_G$, where $r(G)$= rank$(G)$
and $W_G$ is the Weyl group of $G$, with real scalar expectation value
coordinates given by $\Phi ^a_i$, where $a=1\dots 5$ is an index in
the ${\bf 5}$ of $SO(5)_R$ and $i=1\dots r(G)$.

The theory is interacting at the origin and, more generally, at the
boundaries of $\M$, where $\M$ is singular.  On the other hand, for
the generic vacuum in the bulk of $\M$, the massless spectrum is that
of $r(G)$ free, 6d, \N20\ tensor multiplets.  Naively, any effects
associated with degrees of freedom which were massless in the
interacting theory, but become massive for the generic vacuum in the
bulk of $\M$, would decouple at energy scales much less than their
mass, which can be made arbitrarily large by going to large vevs in
$\M$.  One such degree of freedom are BPS strings, which couple to the
$r$ two-form gauge fields with charges $\alpha ^i$, $i=1\dots r$.
These charge vectors span the root lattice of $G$, and the string with
charges $\alpha ^i$ has tension $|\alpha ^i\Phi _i|$ (here $|\Phi
_i|\equiv \sqrt{\sum _a\Phi _i^a\Phi _i ^a}$, the length of the
$SO(5)$ vector), which can be made arbitrarily large by taking the
scalar expectation values $\Phi _i^a$ to be huge.  In the realization
via M5 branes, with separations $\Phi _i^a$ in the 11d bulk, these
strings come from M2 branes which stretch between the M5 branes. Upon
$S^1$ compactification, the W-bosons of $G$ come from these strings
wrapped on $S^1$ \wcomments.

However, no matter how far the vacuum is from the origin of the moduli
space, there are effects associated with the interacting theory at the
origin which can not decouple from the low energy theory.  The reason
is that the interacting theory at the origin generally has a
non-trivial 't Hooft anomaly associated with the global $SO(5)_R$
symmetry.  This anomaly differs from that of the $r$ tensor
multiplets comprising the massless spectrum away from the origin. We
will argue that a Wess-Zumino term must be present in the low energy
theory to account for what would otherwise be a deficit in the 't
Hooft anomaly.

As mentioned above, everything which is presently known about the
interacting, 6d, \N20\ field theory has been obtained from string
theory. (A hope is that it will eventually be understood how to
properly formulate these theories, and recover the properties
predicted via string theory, directly in the context of some sort of
quantum field theory.)  In particular, the non-trivial $SO(5)_R$ 't
Hooft anomaly mentioned above was found in \HMM\ in the context of 11d
$M$ theory, which gave the anomaly for the case $G=SU(N)$, realized as
$N$ parallel M5 branes.  The interesting anomaly coefficient for the
$G=SU(N)$ case was found \HMM\ to be $c(SU(N))=N^3-N$.  The
generalization for the $G=SO(2N)$ and $G=E_{6,7,8}$ cases has not yet
appeared in the literature.

In the next section, we discuss anomaly matching and the
Hopf-Wess-Zumino term which it requires.  In particular, the
coefficient of this term is the difference between the anomaly $c(G)$
of the interacting theory at the origin and that of the low energy
theory away from the origin.  Nontrivial maps in $\pi _7(S^4)$ imply a
non-trivial quantization condition on the WZ term (which is related to
the Hopf invariant of the map) and, consequently, on the anomaly:
$c(G)\in 6\IZ$.  Skyrmionic strings associated with $\pi _4(S^4)$ are
also discussed, and it is proposed that they are the $W$-boson
strings.

In sect. 3, we briefly discuss 4d ${\cal N}=4$ theories, 't Hooft
anomaly matching, and the WZ term thus required in the low-energy theory
when away from the origin.  In this case, the WZ term can be derived
by a standard \refs{\GW, \EDEF} 1-loop calculation
\ATKZ.  We also review some math facts concerning the Hopf
invariant and map.

In sect. 4, we review how the $N^3$ dependence of the entropy and Weyl
anomaly, which is related by supersymmetry to the $SO(5)_R$ anomaly
$c(SU(N))$, was originally found
\refs{\KT, \HS}, via $M$ theory on $AdS_7\times S^4$.  We generalize
this argument to $M$ theory on $AdS_7\times X_4$ for general Einstein
space $X_4$, finding the anomaly (in the large $N$ limit) $c(N;
X_4)=N^3/$vol$(\widehat X_4)^2$.  This argument shows that the anomaly
for the \N20\ theory associated with $G=SO(2N)$ is $c(SO(2N))=4N^3 +$
terms lower order in large $N$.

In sect. 5, we discuss how the needed WZ term of the 
\N20\ theory indeed arises in the world-volume of a M5 brane, which
probes $N$ distant M5 branes.  

In sect. 6, we discuss anomaly matching and a Hopf-WZ term in the 2d
${\cal N}=(0,4)$ CFT which arises in the world-volume of strings in
5d.  This occurs via M theory on a Calabi-Yau three-fold, with the 5d
the uncompactified directions and the strings coming from M5 branes
wrapped on a 4-cycle of the Calabi-Yau.  As will be discussed, perhaps
the story of this section is a fantasy, since there is no moduli space
in 2d.

One might expect that, in the context of field theory, it would be
possible to {\it derive} directly the Wess-Zumino term, by some analog
of the 1-loop computation of
\refs{\GW , \EDEF} for integrating out some massive degrees of
freedom.  Turning around our anomaly matching discussion, this would
give a derivation of the anomaly of the interacting theory at the
origin; e.g. the result of \HMM\ could be re-derived and checked
directly in the context of field theory, without having to invoke
M-theory.  A hope is that these issues could lead to a better
understanding of the interacting 6d \N20\ field theory. In sect. 7, we
speculate on deriving the WZ term via integrating out tensionful
strings and on a possible formula for the 't Hooft anomaly for general
$G=A,D,E$ type \N20\ theories: $c(G)=|G|C_2(G)$.

\newsec{Six dimensional, \N20\ effective field theory}

The \N20\ theory associated with arbitrary group $G$ is expected to
have an anomaly of the following general form when coupled to a
background $SO(5)_R$ gauge field 1-form $A$, and in a general
gravitational background:
\eqn\anomgen{I_8(G)=r(G)I_8(1)+c(G)p_2(F)/24,} 
where $p_i$ are the Pontryagin classes for the background $SO(5)_R$
field strength $F$,
\eqn\pii{p_1(F)= {1\over 2} 
({i\over 2\pi})^2 \tr F^2, \qquad 
p_2(F)={1\over 8}({i\over 2\pi})^4((\tr F^2)\wedge (\tr F^2)-
2\tr F^4).}
(Writing the Chern roots of $F/2\pi$ as $\lambda _1$ and $\lambda _2$,
$p_1=\lambda _1^2+\lambda _2^2$ and $p_2=\lambda _1^2\lambda _2^2$.)
$I_8$ is the anomaly polynomial 8-form, which gives the anomaly by the
descent formalism: $I_8=dI_7^{(0)}$, $\delta I_7^{(0)}=dI_6^{(1)}$,
with $I_6^{(1)}$ the anomalous variation of the Lagrangian under a
gauge variation $\delta$. $I_8(1)$ is the anomaly polynomial for a
single, free, \N20\ supermultiplet \refs{\AW,
\Wiv}:
$I_8(1)=(p_2(F)-p_2(R)+
\f{1}{4} (p_1(R)-p_1(F))^2)/48$. The gravitational anomalies, associated
with any non-trivial curvature $R$, appear only in $I_8(1)$.  In
\anomgen, $r(G)$ is the rank of the group $G$ associated with the \N20\
theory and the quantity $c(G)$, which we refer to as the 't Hooft
anomaly of the $SO(5)_R$ flavor symmetry, also depends on $G$.  The
anomaly \pii\ was found via $M$ theory in \HMM, for the case
$G=SU(N)$, with the result that $c(SU(N))=N^3-N$.  The analog for
for other $G$ has not yet appeared in the literature.

The $SO(5)_R$ current is in the same supermultiplet as the
stress-tensor, and thus the 't Hooft anomaly $c(G)$ also enters in a
term in the Weyl anomaly.  The entropy of the \N20\ theory at finite
temperature is also proportional to $c(G)$.  Indeed, the $N^3$
behavior of $c(G=SU(N))$ was first discovered in these two ways, in
the context of $N$ M5 branes in 11d SUGRA \refs{\KT, \HS}.  Viewing
$c(G)$ as a $c$-function, it should decrease in RG flows to the
IR. E.g. compactifying the 6d theory and flowing in the IR to 4d
${\cal N}=4$, this suggests that in all cases
$c_{UV}=c(G)>c_{IR}=|G|\equiv$dim$(G)$.

The result \anomgen\ gives the anomaly at the origin, where the
$SO(5)_R$ global symmetry is unbroken.  Away from the origin,
$SO(5)_R$ is spontaneously broken.  Nevertheless, we argue that the 't
Hooft anomaly of \anomgen\ must be reproduced everywhere on the moduli
space of vacua.  The argument for 't Hooft anomaly matching is same as
the original argument of 't Hooft in 4d \thooft: we could imagine
adding spectator\foot{In the context of M5 branes, the role of these
``spectators'' is played by contributions from the 11d bulk: the
anomaly inflow and the Chern-Simons term contributions of \refs{\FHMM,
\HMM}.}  fields, which remain decoupled from the rest of the dynamics,
to cancel\foot{In 6d, conjugate group representations contribute to
the anomaly polynomial $I_8$ with the same sign.  Massless fermions of
chiralities $(\half, 0)$ and $(0, \half)$ under the $SO(4)\cong SU(2)
\times SU(2)$ little group contribute with 
opposite signs.} the anomalies, allowing the global symmetry to be weakly
gauged.  The Ward identities of the symmetry must then always be
satisfied. Thus, subtracting the constant contribution of the spectators,
the anomalous Ward identities of the original theory must be
independent of any deformations, including the scalar expectation
values. Away from the origin, a Wess-Zumino term, with specific
coefficient, is needed to ensure that this is the case.

For simplicity, we consider the case that the scalar
vacuum expectation values are chosen to be $\Phi ^a_i=
\phi ^a(T)_{ii}$, where $(T)_{ii}$ are the diagonal components
of a generator of the Cartan of $G$ whose little group is $H\times
U(1)\subset G$.  The massless spectrum for $\ev{\phi ^a}\neq 0$ is
that of the \N20\ CFT associated with $H$, along with a single
additional \N20\ multiplet associated with the $U(1)$.  The $\phi ^a$
are the scalars in the \N20\ supermultiplet associated with this
$U(1)$. Naively, for energy $E\ll \sqrt{|\phi |}\equiv (\phi ^a\phi
^a)^{1/4}$, the $H$ and $U(1)$ theories are decoupled and the \N20\
multiplet associated with the $U(1)$ is free.  

However, the $U(1)$ multiplet of the theory on the Coulomb branch is
actually never really free: it must always include a WZ interaction
term.  The WZ term is needed to compensate for what would otherwise be
a difference in the 't Hooft anomaly \anomgen\ between the $G$ theory
at the origin and the massless $H\times U(1)$ \N20\ theory for $|\phi
|\neq 0$:
\eqn\anomdif{I_8(G)-I_8(H\times U(1))={1\over 24}(c(G)-c(H))p_2(F).}

For $\ev{\phi ^a}\neq 0$, the global $SO(5)_R$ symmetry is broken to
$SO(4)_R$ and the configuration space, for fixed
non-zero\foot{Although there is no potential which requires $\ev{\phi
^a}\neq 0$, it is a modulus labeling superselection sectors, so we can
always choose this to be the case by our choice of boundary conditions
at infinity. The requirement that $|\phi|$ be fixed is not essential:
we only need ${\cal M}_c\cong S^4$ topologically, and requiring
$\ev{\phi ^a}\neq 0$ is enough.}  $|\phi|$, is $\M
_c=SO(5)/SO(4)=S^4$, with coordinates $\widehat \phi ^a\equiv \phi
^a/|\phi |$, $a=1\dots 5$.  The needed Wess Zumino term is given by
the following term in the action
\eqn\WZis{S_{WZ}={1\over 6}(c(G)-c(H))\int _{\Sigma _7}\Omega _3 (\widehat 
\phi , A) \wedge
d\Omega _3(\widehat \phi , A)+\dots,} where $\dots$ are terms related by
supersymmetry.  $\Sigma _7$ is a 7 dimensional space, whose boundary
is the 6d spacetime $W_6$ of the \N20\ field theory, $\partial
\Sigma _7=W_6$ (e.g. $\Sigma _7$ could be $AdS_7$).  
$\Omega _3(\widehat \phi , A)$ is a 3-form which is defined as follows.
Consider the 4-form
\eqn\eivis{\eqalign{\eta _4(\widehat \phi, A)&\equiv \half e^\Sigma _4
\equiv {1\over 64\pi ^2}
\epsilon _{a_1\dots a_5}[
(D_{i_1}\widehat \phi )^{a_1}(D_{i_2}\widehat\phi
)^{a_2}(D_{i_3}\widehat \phi )^{a_3} (D_{i_4}\widehat \phi )^{a_4}\cr
&-2 F^{a_1a_2}_{i_1i_2}(D_{i_3}\widehat
\phi )^{a_3} (D_{i_4}\widehat\phi )^{a_4}+
F^{a_1a_2}_{i_1i_2}F^{a_3a_4}_{i_3i_4}]\widehat \phi
^{a_5}dx^{i^1}\wedge
\cdots
\wedge dx^{i_4},}}
with $(D_i\phi )^a\equiv \partial _i \phi ^a-A_i^{ab}\phi ^b$ the
covariant derivative of $\phi ^a$, involving the background $SO(5)_R$
gauge field $A_i^{ab}=-A_i^{ba}$, with $a, b \in {\bf 5}$ of $SO(5)_R$
($F_{ij}^{ab}$ is its field strength).  The $x^i$ are the coordinates
on $\Sigma _7$.  In \eivis, $e_4^\Sigma \equiv \widehat \phi ^*(e_4)$
is the pullback to $\Sigma _7$, via\foot{Note that to construct the
WZ term requires extending $\widehat \phi$: $W_6\rightarrow S^4$ to
$\widehat \phi$: $\Sigma _7
\rightarrow S^4$, which can have an obstruction if the original 
$\widehat \phi$ is in the non-trivial component of $\pi _6(S^4)=\IZ
_2$.}  $\widehat \phi: \Sigma _7\rightarrow S^4$, of the global,
angular, Euler class 4-form $e_4$ which also entered in \FHMM.  The
$\eta _4$ in \eivis\ is normalized so that $\eta _4(\widehat
\phi ,A=0)=\widehat \phi ^*(\omega _4)$, the pullback of the $S^4$
unit volume form, $\int _{S_4}\omega _4=1$.  The form
\eivis\ is closed and, because we take $\Sigma _7$ such that
$H^4(\Sigma _7)$ is trivial, it must be exact, $\eta 
_4(\widehat \phi, A)=d\Omega _3(\widehat \phi, A)$.  This defines the
$\Omega _3$ appearing in \WZis.

The Wess-Zumino term \WZis\ has the desired non-trivial gauge
variation under $SO(5)_R$ gauge transformations.  To see this, we note
that eqn. (2.7) in \BC\ implies that
\eqn\BCimp{d\Omega _3\wedge d\Omega _3\equiv {1\over 4}e_4^\Sigma \wedge 
e_4^\Sigma ={1\over 4}p_2(F)+d\chi,} where $\chi$ is invariant under
$SO(5)_R$ gauge transformations. Writing the left hand side as
$d(\Omega _3\wedge d\Omega _3)$ and $p_2(F)=dp_2^{(0)}(A)$, the
$SO(5)_R$ gauge variation of \BCimp\ implies that
\eqn\WZvar{\delta \int _{\Sigma _7}\Omega _3\wedge d\Omega _3=
{1\over 4}\int _{\Sigma _7}\delta p_2^{(0)}(A)=
{1\over 4}\int _{W_6}p_2^{(1)}(A),}
where $p_2^{(1)}(A)$ is the anomaly 6-form found by descent, $\delta
p_2^{(0)}=dp_2^{(1)}$.  Note that the $\phi$ dependence in $\Omega
_3(\widehat \phi, A)$ has dropped out in the gauge variation \WZvar.
Using \WZvar, the $SO(5)_R$ gauge variation of the WZ
term \WZis\ indeed compensates for the deficit \anomdif.

As an example, consider $G=SU(N+1)$ and $H=SU(N)$.  Using the
result of \HMM\ that $c(G)=(N+1)^3-(N+1)$ and $c(H)=N^3-N$, 
the Wess-Zumino term \WZis\ is 
\eqn\sunwz{\half N(N+1)\int _{\Sigma _7}\Omega _3(\widehat \phi, A)\wedge 
d\Omega _3 (\widehat \phi , A).}  This Wess-Zumino term must be
present in the world-volume of a M5 brane, when in the background of
$N$ other M5 branes, and thus for a M5 brane in $AdS_7\times S^4$.

By a general analysis\foot{I thank E. D'Hoker for pointing this out to
me and for related correspondences.} \EDSW, WZ terms are generally of
the form $\int _{\Sigma _{d+1}}
\widehat \phi ^*(\omega _{d+1})$, with $\omega _{d+1}\in 
H^{d+1}(\M _c, \IR)$.  However, our WZ term \WZis\ is not of this form,
as $\Omega _3 \wedge d\Omega _3\neq \widehat \phi ^*(\omega _7)$.
Indeed, here $\M _c=SO(5)/SO(4)\cong S^4$, and obviously $H^{7}(S^4,\IR
)=0$; nevertheless, even for $A=0$,
\WZis\ is non-zero.  An aspect of the present case, which 
sets it apart from the general analysis of \EDSW, is mentioned at the
end of this section.

The 7-form $\Omega _3\wedge d\Omega _3$ in \WZis\ is not exact, so the
ambiguity in the choice of $\Sigma _7$ is non-trivial.  The difference
between choosing $\Sigma _7$ and $\Sigma _7'$, both with boundary
$W_6$, is the integral over $\Sigma _7-\Sigma _7'\cong S^7$
\eqn\WZdif{{1\over 6}(c(G)-c(H))
\int _{S^7}\Omega _3 (\widehat \phi , A) \wedge
d\Omega _3(\widehat \phi , A).}  This can be non-trivial.  Indeed,
e.g. for zero background $SO(5)_R$ field, $A=0$, the integral in
\WZdif\ gives the Hopf number of the map $\widehat \phi ^a:
S^7\rightarrow S^4$, which can be an arbitrary integer, corresponding
to $\pi _7(S^4)=\IZ +\IZ _{12}$.  The coefficient of \WZdif\ thus must
be quantized in order for $e^{2\pi iS}$ to be well-defined and invariant
under the choice of $\Sigma _7$:
\eqn\cdiffq{{1\over 6}(c(G)-c(H))\in \IZ.}
To have \cdiffq\ hold for arbitrary $ADE$ groups $G$ and subgroups $H$
requires
\eqn\cquant{{1\over 6}c(G)\in \IZ}
for all $ADE$ groups $G$.  Happily, \cquant\ is indeed satisfied by
$c(G=SU(N))=N^3-N$.

We also note that there are topologically stable, solitonic
``skyrmion'' field configurations in the theory with non-zero
$|\phi|$.  In $d$ spacetime dimensions, a $p$-brane skyrmion is a
field configuration $\widehat \phi ^a(X_t)$ which only depends on the
$d-p-1$ space coordinates of $X_t$, the space transverse to the
$p$-brane worldvolume.  In order for this to be a finite-energy
configuration, $\widehat \phi ^a$ must approach a constant value when
the coordinates of $X_t$ are taken to infinity.  Such field
configurations are thus topologically classified by $\pi _{d-p-1}(\M
_c)$.  In the present case, $\pi _4(S^4)=\IZ$ means that there are
non-trivial $p=1$ branes in $d=6$, i.e. there are skyrmionic strings.
(There are also $\IZ _2$ particles since $\pi _5(S^4)=\IZ _2$.)  The
topological charge density for the skyrmionic strings is $\eta _4$,
defined as in \eivis: the string number is $N_s=\int _{X_t}
\eta _4$.  The WZ term means that there is a Goldstone-Wilczek
contribution
\refs{\GW, \EWWZ} to the $SO(5)_R$ flavor current, which can give the
skyrmions $SO(5)_R$ charges.


We propose that these skyrmionic strings are actually the ``W-boson''
BPS strings mentioned in the introduction.  (Other works, including
\refs{\OGLM, \Getal}, have 
briefly considered solitonic strings in the M5 brane theory, but not
specifically the $\pi _4(S^4)$ skyrmionic solitons.)  In line with
this proposal, the skyrmionic string density $\eta _4$ should act as
electric and magnetic flux sources for the $H_3$ in the $U(1)$ \N20\
multiplet:
\eqn\skyhf{dH_3=J_{mag}=\alpha _m\eta _4,\qquad 
d*H_3=J_{elec}=\alpha _e \eta _4} for some non-zero constants $\alpha
_{m,e}$.  The $H_3$ in \skyhf\ is not self-dual, rather it is related
to a self-dual tensor $h_3$ by a non-linear transformation \HSW, so
$\alpha _e$ and $\alpha _m$ need not be equal.  The electric relation
in \skyhf\ means there is an interaction
\eqn\BScoup{S_{sky}=\alpha _e \int _{W_6}B_2\wedge \eta _4=
-\alpha _e\int _{W_6}dB_2\wedge \Omega _3(\widehat \phi ,A).}

Given that the skyrmionic string is charged under $H_3$ as outlined
above, it follows from completely general considerations that the
supersymmetry algebra has central term $Z=|Q\phi |$, and the tension
of such a string satisfies $T\geq |Q\phi |$.  Here $Q\sim N_s$, with
$N_s$ the $\pi _4(S^4)$ topological string number, $N_s=\int
_{X_t}\eta$. For each $N_s$ charge, there is a BPS field configuration
$\widehat \phi ^a$ which minimizes the energy, satisfying $T=|Q\phi|$.
It is these BPS skyrmionic string solitons which should be identified
with the BPS W-boson strings.

In the particular context of $N$ M5 branes, corresponding to the $G=SU(N)$
\N20\ theories, the magnetic relation in \skyhf\ and the 
coupling \BScoup\ have already appeared in \OGLM, though the
interpretation of these relations in terms of the $\pi _4(S^4)$
skyrmionic strings was not discussed there.  The argument of \OGLM\
for the magnetic relation in \skyhf\ involves accounting for the fact
that M5-branes act as $G_4$ sources; this is also related to the
analysis and results of \refs{\FHMM, \HMM}, and is further discussed
in sect. 5. 

The argument of \OGLM\ for the coupling \BScoup, which immediately
generalizes to general \N20\ $G\rightarrow H\times U(1)$ Coulomb
branch Higgsing, is as follows (see also \refs{\CBGT,
\ATKZ}): consider $S^1$ reducing to 5d, where the
theory is ordinary Yang-Mills and is IR free.  The 5d $U(1)$ gauge
field in the Higgsing $G\rightarrow H\times U(1)$ arises from $B_{\mu
6}$ in 6d.  The coupling
\BScoup\ then arises by a standard type of 1-loop calculation, much
as in \refs{\GW, \EDEF}, with the $n_W=|G|-|H|-1$ massive gauginos
running in the loop.  Taking care with the normalization, we get 
\BScoup\ with $\alpha _e={1\over 4} n_W$. 
E.g. for $G=SU(N+1)$ and $H=SU(N)$, the term
\BScoup\ is generated with coefficient $\half N$.
The constant $\alpha _m$ in \skyhf\ is more difficult to determine.
 
We can also consider $S^1$ dimensional reduction of the relations
\skyhf, using $H_3\rightarrow H_3+F_2\wedge dx_6$ and $\eta
_4\rightarrow \eta _4$, where now, using \eivis, $\eta _4$ has no
$dx_6$ component because we take $\widehat \phi$ to be independent of
$x_6\in S^1$ in dimensional reduction. Then \skyhf\ gives $dH_3=\alpha
_m \eta _4$, $dF_2=0$, $d*H_3=0$, and $d*F_2=\alpha _e
\eta _4$ (now $*$ acts in the uncompactified 5d).  The $F_2$ equations
show that there are no magnetic charges in 5d, and that the $\pi
_4(\M _c=S^4)$ skyrmions are electrically charged particles in 5d which,
since $\alpha _e={1\over 4}n_W$, can be identified with the $n_W$
electrically charge $W$-bosons.  Naively, one might identify $H_3$ as
$*F_2$, making the $H_3$ equations repeats of the $F_2$ equations and
suggesting that $\alpha _m$ be identified with $\alpha _e$.  However,
as inherited from 6d where $H_3$ is not simply self-dual, the 5d $H_3$
and $*F_2$ are not simply equal. This again makes $\alpha _m$ more
difficult to determine.

We emphasize that, unlike the term \BScoup, it does not seem possible
(at least in any obvious way) to get our 6d Hopf-Wess-Zumino term
\WZis\ by a direct calculation in the dimensionally reduced 5d gauge theory. 
Indeed, upon $S^1$ dimensional reduction, the term \WZis\ actually
vanishes (unless the Kaluza-Klein $S^1$ momentum modes are included)
since, for $\widehat \phi$ independent of $x_6$, \eivis\ shows that
$\eta _4$, and thus also $\Omega _3$, have no $dx_6$ component.  It
could have been anticipated that the 6d WZ term \WZis\ would be
difficult to obtain by dimensional reduction because (unlike \BScoup)
its coefficient is not simply $n_W$; e.g. in
\sunwz\ the coefficient is proportional to $N(N+1)$ rather than just
$n_W\sim N$.

Here is a possible insight into the origin of the 6d WZ term \WZis: in
direct analogy with the discussion in \EWWZ, the action of an electric
current in a magnetic background is $\int _{\Sigma _7}H_3 \wedge
J_{elec}$, again with $\partial
\Sigma _7=W_6$.  Solving
\skyhf\ for $H_3$ as \eqn\His{H_3=dB_2+\alpha _m\Omega _3(\widehat \phi ,A),}
with $\eta _4=d\Omega _3$, and plugging in $J_{elec}$ from \skyhf,
this gives the action
\eqn\seandm{\int _{\Sigma _7}\alpha _e(dB_2+\alpha _m\Omega _3)\wedge 
d\Omega _3.}  The first term in \seandm\ gives the coupling
\BScoup\ and the second term gives the WZ term \WZis, with the right
coefficient (assuming that the WZ term indeed arises entirely from
\seandm) provided that ${1\over 6}(c(G)-c(H))=\alpha _e\alpha
_m={1\over 4}n_W\alpha _m$. In this light, \cdiffq\ is simply Dirac
quantization.  Note that, in $\Sigma _7$, $\eta _4$ becomes a density
for skyrmionic membranes, whose ends on $\partial
\Sigma _7=W_6$ are the skyrmionic strings of $W_6$ discussed above. 
In the M5 brane realization, these are like skyrmionic M2 branes
living, e.g. in $\Sigma _7=AdS_7$.  The WZ term is proportional to
$\int _{\Sigma _7} \Omega _3\wedge d\Omega _3$, which measures
membrane winding number in $\Sigma _7$.

Lastly, we tie up a loose end: our definition of $\Omega _3$, via
$\eta _4=d\Omega _3$, only defines $\Omega _3$ up to exact forms,
$\Omega _3\rightarrow
\Omega _3+d\Lambda _2$.  Under such a change, \WZis\ 
changes by
\eqn\WZchange{S_{WZ}\rightarrow S_{WZ}-{1\over 6}(c(G)-c(H))\int
_{W_6} d\Lambda _2\wedge \Omega _3;} this freedom must somehow be
fixed in order for the effective action to be well-defined\foot{I
thank E. Witten for correspondences, which stressed the need to fix
this issue and suggested the following discussion.}.  Noting that the
physical quantity $H_3$ must also be well-defined, \His\ shows that
the change $\Omega _3\rightarrow \Omega _3+d\Lambda _2$ requires a
compensating shift $B_2\rightarrow B_2-\alpha _m
\Lambda _2$.  (In the M5-brane realization, to be discussed in
sect. 5, this is the freedom of $C_3$ gauge transformations.)  If the
WZ term arises from \seandm, it is unchanged by this combined shift of
$\Omega _3$ and $B_2$, with the change in \BScoup\ under
$B_2\rightarrow B_2-\alpha _m
\Lambda _2$ cancelling \WZchange.  Alternatively, we can simply use
\His\ to define $\Omega _3$ on $W_6$ as $\alpha _m^{-1}H_3$. The
remaining ambiguity on $\Sigma _7$ of taking $\Omega _3\rightarrow
\Omega _3+d\Lambda _2$, with $d\Lambda _2|_{W_6}=0$, is harmless in
\WZchange.  In short, our WZ term \WZis\ needs $B_2$ to be 
well-defined, an aspect which sets it apart from the general analysis
of \EDSW.

\newsec{Miscellaneous Notes}

\subsec{The WZ Term of 4d ${\cal N}=4$ Theories}

A completely analogous relation between 't Hooft anomaly matching and
a WZ term holds in the 4d ${\cal N}=4$ theory.  The ${\cal N}=4$
theory has a global $SU(4)_R\cong SO(6)_R$ flavor symmetry, with 't
Hooft anomaly $\tr SU(4)_R^3=|G|$, i.e. in a background $SU(4)_R$
gauge field $A_B$, with field strength $F_B$, there is an anomaly
determined via descent from
\eqn\ivdt{
I_6(G)={|G|\over 6}({i\over 2\pi})^3\tr F_B^3,} with $|G|$ the dimension
of the gauge group $G$.  This anomaly comes from the $|G|$ gauginos in
the ${\bf 4}$ of $SU(4)_R$ and is not renormalized.

Consider now moving away from the origin of the moduli space via $\Phi
^a= \phi ^aT$, with $T$ a generator of the Cartan of $G$, with little
group $H\times U(1)\subset G$.  Here $a\in {\bf 6}$ of $SU(4)_R$ and
taking $\ev{\phi ^a}\neq 0$ breaks the gauge symmetry $G\rightarrow
H\times U(1)$ and the flavor symmetry $SU(4)\rightarrow SO(5)$.  For
fixed $|\phi|\equiv \sqrt{\phi ^a\phi ^a}$, the configuration space is
$\M _c=SU(4)/SO(5)\cong S^5$.  The massless spectrum is that of the
${\cal N}=4$ theory with decoupled groups $H\times U(1)$ and, for
energies $E\ll |\phi|$, one might be tempted to forget about the
effects of the $n_W=|G|-|H|-1$, ultra-massive,
$G/H\times U(1)$ gauge field multiplets.  
However, the $n_W$ gauginos in these multiplets contributed to the
anomaly \ivdt; without them there is a deficit in \ivdt\ of 
$I_6(G)-I_6(H)-I_6(U(1))={1\over 6}n_W({i\over 2\pi})^3 \tr F_B^3$.
This deficit must be accounted for by a Wess-Zumino term in the low
energy theory.

The WZ term thus required in the low-energy theory is
\eqn\ivWZis{\half n_W \Gamma [\widehat \phi , A_B]+
\hbox{superpartners},}
where $\Gamma [\widehat \phi, A_B]$ is conventionally \EWWZ\ written
as $\Gamma [\widehat \phi, A_B] =\Gamma [\widehat \phi]+Z[\widehat
\phi, A_B]/48\pi ^2$ with
\eqn\Gammais{\Gamma [\widehat \phi]={1\over 240\pi ^2}
\int _{\Sigma _5}\epsilon _{a_1 \dots a_6}\partial _{i_1}\widehat \phi ^{a_1} 
\dots \partial _{i_5}\widehat \phi ^{a_5}\widehat \phi ^{a_6}dx^{i_1}
\wedge \dots \wedge dx^{i_5},}
where $\partial \Sigma _5=W_4$.  $\Gamma [\widehat \phi,A]=\half \int
_{\Sigma _5}\widehat \phi ^*(e_5)$, with $e_5$ the $S^5$ global,
angular, Euler class form in the appendix of \HMM, normalized so that
$\half e_5(A=0)=\omega _5$, the unit $S^5$ volume form. (Since $\pi
_4(S^5)=0$, there is no obstruction to extending $\widehat \phi:
W_4\rightarrow S^5$ to $\widehat \phi: \Sigma _5\rightarrow S^5$.)
Corresponding to
\ivWZis, there is an induced Goldstone-Wilczek current
\eqn\icGWc{j^\mu _{a_1a_2}=\half n_W{1
\over 24\pi ^2}\epsilon^{\mu \nu \rho \sigma}\epsilon _{a_1a_2 \dots
a_6}\partial _\nu \widehat \phi ^{a_3}\partial _\rho \widehat 
\phi ^{a_4}\partial _\sigma \widehat \phi ^{a_5} \widehat \phi ^{a_6}.}

The same $\Gamma [\widehat \phi, A_B]$ appeared in \manohar\ in the
context of ${\cal N}=1$ SUSY QCD with $N_f=N_c=2$, as in both cases
there is a $SU(4)$ flavor symmetry with order parameter in the ${\bf
6}$ of constant magnitude.  As shown in \manohar, the $SU(4)$
variation of $\Gamma [\widehat \phi, A_B]$ contributes to the $SU(4)^3$
flavor 't Hooft anomalies the same as with {\it two} fermions in the ${\bf
4}$ of $SU(4)$.  Thus, with the coefficient of the WZ term as in
\ivWZis, it properly accounts for the contribution to the 't Hooft
anomaly of the $n_W$ gauginos, in the ${\bf 4}$ of $SU(4)$, which got
a mass via Yukawa couplings to $\phi$ in the Higgsing $G\rightarrow
H\times U(1)$.  The fact that integrating out the $n_W$ massive
fermions actually does generate precisely the WZ term \ivWZis\
follows from the standard 1-loop calculation of the type appearing in
\refs{\GW, \EDEF}.  See, in particular, \ATKZ.

Because the WZ 5-form term in $\Gamma [\widehat \phi, A_B]$ is not
exact, there is a quantization condition on its coefficient \ivWZis\ in
order to have $e^{2\pi iS}$ be invariant under $\Sigma _5\rightarrow \Sigma
_5'$ with $\partial \Sigma _5' =\partial \Sigma _5=W_4$.  The difference
involves the 5-form of \Gammais\ integrated over $\Sigma _5'-\Sigma _5
\cong S^5$, which is an arbitrary integer associated with 
$\pi _5(\M _c=S^5)=\IZ$.  The quantization condition is thus
\eqn\ivdq{\half n_W\equiv \half (|G|-|H|-1)\in \IZ.}
Fortunately, this is indeed satisfied for arbitrary group $G$, with
subgroup $H\times U(1)$ obtained via adjoint Higgsing.  Since all $\pi
_{3-p}(\M _c=S^5)=0$, now there are no $p$-brane skyrmions.

The 4d WZ term \ivWZis\ is related to the dimensional reduction of
\BScoup, not the 6d WZ term \WZis.  Again, the dimensional reduction
of the 6d WZ term vanishes.

\subsec{Some math notes on the Hopf invariant}
We now summarize some facts which can be found e.g. in \Hilton.  The
Hopf invariant $H(f)$ of a mapping $f: S^{2n-1}\rightarrow S^n$ is an
integer which can be defined as the winding coefficient of curves
$f^{*}(a)$ and $f^{*}(b)$ in $S^{2n-1}$ for distinct $a$ and $b$ in
$S^n$; $f^{*}(a)$ is the $(n-1)$ dimensional curve in $S^{2n-1}$ which
is mapped by $f$ to a point $a\in S^n$.  $H(f)$ can be written as an
integral over $S^{2n-1}$ as follows: consider the pullback
$f^{*}(\omega _n)$, where $\omega _n$ is the unit volume form of
$S^n$, $\int _{S^n}\omega _n=1$.  The form $f^*(\omega _n)$ is closed
and, as $H^n(S^{2n-1})$ is trivial, must be exact, $f^{*}(\omega
_n)=d\theta _{n-1}$. The Hopf invariant can be written as
\eqn\Hfis{H(f)=\int _{S^{2n-1}}\theta _{n-1}\wedge d\theta _{n-1}.}
Clearly, $H(f)=0$ for $n$ odd.  For $n=2k$ even, $H(f)\in
\IZ$, taking all integer values for various maps $f$. Thus 
$\pi _{4k-1}(S^{2k})$ is at least $\IZ$.  E.g. $\pi _3(S^2)=\IZ$ and
$\pi _4(S^7)=\IZ \oplus \IZ _{12}$.

The basic map $S^3\rightarrow S^2$ with Hopf number 1 is given 
by writing $S^3$ as $(z_1,z_2)$, with $z_i\in \IC$ and 
$|z_1|^2+|z_2|^2=1$, and writing $S^2$ as $CP^1$, i.e. 
$[z_1,z_2]$ with
$z_i\in \IC ^*$ and $[z_1,z_2]\sim [\lambda z_1, \lambda z_2]$
for arbitrary $\lambda \in \IC ^*$.  The map is then simply
$f: (z_1,z_2)\rightarrow [z_1, z_2]$.  The map with Hopf number
1 for $S^7\rightarrow S^4$ is exactly the same as that above,
with the simple replacement that $z_i$ and $\lambda$ now take
values in the quaternionic rather than the complex numbers.

\newsec{Getting 
the $N^3$ via gravity and the $G=SO(2N)$ case via an orbifold} 

We now review how $c\approx N^3$ appears via 11d sugra, generalizing
to $M$ theory on $AdS_7\times X_4$, where $X_4$ is a general, compact,
Einstein space.  The anomaly coefficient $c$ arises as the coefficient
of a Chern-Simons term in $AdS_7$.  This term is related by
supersymmetry to the coefficient of the 7d Einstein-Hilbert action in
$AdS_7$.  It thus follows that
\eqn\cis{c={L^5\over G_7},}
where $G_7$ is the 7d Newtons constant, and the powers of $L$, which
is the horizon size of $AdS_7$ (related to the size of the negative
cosmological constant), are determined by dimensional analysis; for
simplicity, we will everywhere drop universal constants (factors of 2
and $\pi$).  The entropy \KT\ and Weyl anomaly \HS\ are also
proportional to \cis.

By the dimensional reduction from 11d SUGRA or M theory on compact
space $X_4$, $G_7^{-1}=$vol$(X_4)/l_P^9$, with $l_P$ the 11d Planck
length.  We thus write \cis\ as
\eqn\ciss{c={\rm vol}(\widehat X_4){L^9\over l_P^9},}
where vol($\widehat X_4$) is the dimensionless volume of $X_4$
measured in units of $L$ (normalized so that vol$(\widehat X_4)=1$ for
$X_4=S^4$ of radius $L$).  The $G_4$ flux quantization condition gives
\eqn\gflux{\int _{X_4}G_4 =L^3{\rm vol}(\widehat X_4)=Nl_P^3,}
so \ciss\ leads to the general result
\eqn\cfin{c={N^3\over ({\rm vol} (\widehat X_4))^2}+\hbox{lower
order in }N.}
In particular, for orbifolds $X_4=S^4/\Gamma$, \cfin\ gives
\eqn\corbis{c=N^3|\Gamma |^2+\hbox{lower order in }N,}
in the normalization where $c=N^3$ (plus lower order) for the \N20\
theory with $G=SU(N)$, corresponding to $X_4=S^4$.  This argument is
analogous to that of \Gubser\ for IIB on $AdS_5\times X_5$, which gave
$c=N^2/{\rm vol}(\widehat X_5)=N^2|\Gamma|$.

In particular, the \N20\ theory with group $G=SO(2N)$ arises from
$M$ theory on $AdS_7\times RP^4$ and, writing $RP^4=S^4/\Gamma$ 
with $\Gamma = \IZ _2$, \corbis\ implies that the anomaly is
\eqn\cso{c(G=SO(2N))=4N^3+\hbox{lower order in }N.}

\newsec{The WZ term via the M5-brane worldvolume action}

Branes in string or M theory always have some sort of ``Wess-Zumino''
terms, e.g. for $Dp$ branes it is usually written as \DWZR
\eqn\DWZis{S_{WZ}=\int _{W_{p+1}}C\wedge \tr \exp(i(F-B)/2\pi)\wedge
\sqrt{{\widehat A(R_T)\over \widehat A (F_N)}},}
and the presence of some similar terms for the M5 brane is well-known
\OH.  As written, these could
not be exactly the Wess-Zumino terms of the type we have argued for,
as they are written as local integrals over the world-volume $W$ and
not over a higher dimensional space $\Sigma$ with $\partial \Sigma=W$.
Of course, they could be written as an integral over $\Sigma$ of an
exact form, but our WZ term is the integral over $\Sigma$ of a form
which is {\it not} exact.  Nevertheless, we argue that writing the
``Wess-Zumino'' term of \OH\ as the integral over $\Sigma _7$ of a
7-form, which is naively exact, actually gives the Hopf-Wess-Zumino
term which we want.  The point is that the naively exact 7-form
actually is {\it not} exact upon properly taking into account the fact
that 5-branes act as a non-trivial source for $G_4$ in M theory.

Similarly, for Dp-branes, the WZ term generally can {\it not} be
written as the local term \DWZis\ on $W_{p+1}$.  It must be written as
$\int _{\Sigma _{p+2}}\Omega _{p+2}$, with $\Omega _{p+2}$ not exact,
despite the fact that, naively, $\Omega _{p+2}=d\Omega _{p+1}$, with
$\Omega _{p+1}$ the form in
\DWZis.  E.g. for a D3 brane \DWZis\ contains $\int _{W_4}C_4$,
which should really be written as $\int _{\Sigma _5} F_5$. Naively
$F_5=dC_4$ and there is no difference; however, in the presence of
other D3 brane sources, $F_5$ is not exact.  This is how \ivWZis\
arises for a D3 brane probing other D3 branes.

The M5 brane world-volume theory depends on (with sign conventions of 
\refs{\Wiv, \HMM})  
\eqn\htilde{H_3=dB_2+C^W_3,}
with $B_2$ the two-form gauge field and $C^W_3$ the pull-back of the
11d $C_3$ field to the M5 brane world-volume $W_6$.  This $H_3$ 
\htilde\ is invariant under the gauge invariance $\delta C_3=d\Lambda
_2$, $\delta B_2=-\Lambda ^W_2$ and satisfies a generalized
self-duality condition (it is only self-dual at linear order; there is
a field transform to a 3-form $h$ which is exactly self-dual \HSW).

We consider a probe brane in the background of the $N$ others; for
large $N$, this should be equivalent to a M5 brane in $AdS_7\times
S^4$.  Following \HMM, there is a $G_4$ background, with pullback
$G^W_4=N\eta _4(\widehat
\phi, A)+$ fluctuations in $C_3$, with $\eta _4$ the 4-form \eivis;
thus 
\eqn\Ciss{C^W 
_3=N\Omega _3(\widehat \phi , A) +\hbox{fluctuations},} and
\htilde\ becomes \His\ with $\alpha _m\approx N$. The fact 
that $dH_3 \approx N\eta _4$, which follows from 
\htilde\ and \Ciss,  has already been suggested (with $A_{SO(5)}=0$)
in \OGLM.

We re-write the WZ term of \OH\ as an integral over some $\Sigma _7$
with $\partial \Sigma _7 = W_6$:
\eqn\mvwz{\int _{\Sigma _7}(*G_4+\half (dB_2+C^\Sigma_3)\wedge
G^\Sigma _4).} Plugging in $C_3^\Sigma$ and $G_4^\Sigma =dC_3^\Sigma$,
given by \Ciss\ extended to $\Sigma _7$, we get
\eqn\mvwzz{\int _{\Sigma _7}(*G_4 +\half N(dB_2+ 
N\Omega _3) \wedge d\Omega _3).}  
This indeed contains the Hopf-WZ term
\sunwz, with the correct leading order in large $N$ coefficient of 
$\f{1}{2}N^2$.  Indeed, ignoring the $*G_4$ term, \mvwzz\ is of
exactly the form \seandm\ with $\alpha _e\approx \half N$ and $\alpha
_m\approx N$; so \mvwzz\ also contains the coupling \BScoup\ needed
for the $\pi _4(S^4)$ solitonic strings to couple electrically to the
$B_2$ field as the $n_W=2N$ ``W-boson'' strings.  As discussed in
sect. 2, $S^1$ dimensional reduction suggests that we get the term
\BScoup\ with $\alpha _e={1\over 4}n_W=\half N$ (exact).

We should also get the term proportional to $N$ in
\sunwz.  The term $*G_4$ in \mvwz\ will be order $N$, but $*G_4$ needs to
be properly interpreted to see if it also contributes to the
Hopf-Wess-Zumino term (naively it's just a contribution to the $AdS_7$
vacuum energy).  Perhaps a new term, similar to the $C_3\wedge
I_8^{inf}$ term of 11d SUGRA, is needed to get the order $N$ term in
\sunwz.  If the WZ term indeed arises entirely as in \seandm, with
coefficient $\alpha _e\alpha _m$, the order $N$ term in \sunwz\
should arise from correcting $\alpha _m\approx N$ to $\alpha _m=N+1$ .

\newsec{Reduction on a Calabi-Yau 3-fold}

Following the discussion in \refs{\FHMM, \HMM}, we now consider $M$
theory on a Calabi-Yau 3-fold $X$, with M5 branes wrapping a
four-cycle to yield strings.  These strings live in the 5
uncompactified dimensions of $M$ theory on $X$ and their world-volume
theory is a 2d ${\cal N}=(0,4)$ CFT, which has a $SO(3)_R$ global
symmetry.  The $SO(3)_R$ symmetry is that of the normal bundle of the
three transverse directions of these strings in 5d.  E.g. there are 2d
world-volume scalars $\Phi ^a$, $a=1,2,3$, in the ${\bf 3}$ of
$SO(3)_R$, whose expectation values gives the positions of the strings
in the 3 transverse directions.

Classically, we can consider the situation of separating one string
{}from $N$ others in these three transverse directions.  This would
spontaneously break the $SO(3)_R$ global symmetry of the probe string
world-volume theory to an $SO(2)_R$ subgroup.  However, this can not
really happen: there is no spontaneous symmetry breaking in 2d
\coleman.  There is no moduli space, as the scalars $\Phi ^a$ have a
wavefunction which spreads over all values.

We now discuss how the story with anomalies and the WZ term would go
if we ignore the fact that there is actually no moduli space in 2d.
Perhaps this discussion is relevant in some sort of Born-Oppenheimer
approximation, where the spreading of the wave-function is initially
neglected or suppressed.  Or perhaps this section is just a fairy
tale.

The $SO(3)_R$ symmetry is an affine Lie algebra, with level $k$, which
is the $SO(3)_R$ 't Hooft anomaly in the 2d anomaly polynomial
\eqn\Iivis{I_4(G,X)={1\over 4}k(G,X)p_1(F),}
with $F$ the $SO(3)$ background field and $p_1(F)$ as in \pii. (We
ignore gravitational contributions to $I_4$ since they do not
require a WZ term.)  We expect that $G$ can be $U(1)$ or an
ADE group, corresponding to the ADE classification
of $SU(2)_k$ modular invariant partition functions. For $G=U(N)$ 
\refs{\HMM, \MSW}
\eqn\kis{k(U(N),X)=N^3D_0+Nc_2\cdot P_0/12,}
where $D_0$ and $c_2\cdot P_0$ are determined in terms of the geometry
of the 3-fold $X$ and the 4-cycle of $X$ on which the $N$ M5 branes
wrap.

In the fairy tale where we can consider fixed non-zero
$|\phi|=\sqrt{\phi ^a\phi ^a}$, the classical configuration space is
$\M _c=SO(3)/SO(2)=S^2$ and $G$ is broken to $H\times U(1)$.  There
must be a Wess-Zumino term on the probe string world-volume to
compensate for the deficit $I_4(G,X)-I_4(H,X)-I_4(U(1),X)$. 

We take the string world-volume to be $W_2=\partial \Sigma _3$.
Consider the two-form,
\eqn\eiiis{\eta _2(\widehat \phi ,A)\equiv \half e^\Sigma _2=
{1\over 8\pi}\epsilon_{abc}(D_i\widehat \phi ^aD_j\widehat
\phi ^b-F_{ij}^{ab})\widehat \phi ^cdx^i\wedge dx^j,}
where the covariant derivatives include a background $SO(3)$ gauge
field $A^{ab}=-A^{ba}$, with background field strength $F^{ab}$, and
$\widehat
\phi ^a= \phi ^a/|\phi |$.  $e_2^\Sigma$ is the pullback (via\foot{
There can be an obstruction to extending $\widehat \phi$ from $W_2$ to
$\Sigma _3$ if $\widehat \phi$ is in the non-trivial component of $\pi
_2(\M _c=S^2)=\IZ$.}  $\widehat \phi :\Sigma _3\rightarrow S^2$) to
$\Sigma _3$ of the global, angular, Euler-class form $e_2$ appearing
in \FHMM.  Because $H^2(\Sigma _3)$ is trivial, the form \eiiis\ must
be exact, $\eta _2 =d\Omega _1(\widehat \phi ,A)$. The
Hopf-Wess-Zumino term is
\eqn\iidwz{ \Gamma =(k(G,X)-k(H,X)-k(U(1),X))
\int _{\Sigma _3}\Omega _1(\widehat 
\phi, A)\wedge d \Omega _1 (\widehat \phi, A).}

To see that \iidwz\ contributes to the anomaly matching, we note that
\BC
\eqn\eiisq{d\Omega _1\wedge d\Omega _1 \equiv {1\over 4}
e^\Sigma _2(\widehat \phi, A)\wedge e^\Sigma _2(\widehat \phi,
A)={1\over 4}p_1(F)+d\chi,} where $p_1(F)$ is as in \pii\ and $\chi$ is
invariant under $SO(3)$ gauge transformations.  Writing 
$p_1=dp_1^{(0)}$, with $\delta p_1^{(0)}=dp_1^{(1)}$,
\eiisq\ implies 
\eqn\wzvar{\delta \int _{\Sigma _3}\Omega _1\wedge d\Omega _1=
{1\over 4}\int _{\Sigma _3}\delta p_1^{(0)}={1\over 4}\int
_{W_2}p_1^{(1)},} so \iidwz\ compensates for the deficit in the
anomaly \Iivis\ in the low energy theory.

The coefficient of the WZ term must be quantized and properly
normalized in order for $e^{2\pi i \Gamma}$ to be invariant
under changing $\Sigma \rightarrow \Sigma '$, with $\partial
\Sigma = \partial \Sigma '=W_2$.  The difference is \iidwz\
integrated over $S^3\cong \Sigma - \Sigma '$, which  
gives $(k(G,X)-k(H,X)-k(U(1),X))H[\widehat \phi]$, where $H[\widehat 
\phi]$ is the Hopf number of the map $\widehat \phi: S^3
\rightarrow S^2$;  $H[\widehat \phi ]\in \IZ$, 
corresponding to $\pi _3(S^2)=\IZ$. To have the functional integral 
be well defined under $\Sigma \rightarrow \Sigma '$ thus
requires all $(k(G,X)-k(H,X)-k(U(1),X))$, and thus all
$k(G,X)$, to always be an integer.

\newsec{Speculations}

Because the WZ term \WZis\ of the \N20\ theory is related to an
anomaly, it is natural to expect that it can be found exactly by a
1-loop calculation, with some fields which become massive due to
$\ev{\phi}
\neq 0$ running in the loop.  In the 4d ${\cal N}=4$ theory, these
were the $n_W=|G|-|H|-1$ gauginos, which get a mass via Yukawa
couplings to $\ev{\phi}\neq 0$.  The analog in the 6d \N20\ theory are
the BPS strings, which get a tension $1/\alpha '(\phi )\sim \phi$.
Perhaps, then, it is possible to derive the WZ term directly by a
1-loop string calculation, with these $n_W$ strings, coupling to
$\phi$, running in the loop.  This suggests a WZ term proportional to
$n_W$, though we know from \sunwz\ that it can not be exactly just
$n_W$.  Indeed, following \seandm, we speculated that the WZ
coefficient is
\eqn\WZcm{{1\over 6}(c(G)-c(H))=\alpha _e \alpha _m=
{1\over 4}n_W\alpha _m,} e.g. with $\alpha _m=N+1$ for the case
\sunwz. So then the challenge is to get the factor of $\alpha _m$.

We have not yet demonstrated that such a derivation of the WZ term
\WZis, via integrating out tensionful strings, is actually possible.
One might object that the \N20\ theory is really a field theory, and
the strings are not fundamental but, rather, some kind of solitonic
objects, e.g. the skyrmionic strings of sect. 2.  Perhaps, then,
these are not the correct degrees of freedom to be integrating out in
deriving the WZ term.  On the other hand, perhaps the distinction
between fundamental vs composite degrees of freedom is irrelevant for
deriving the WZ term, since it is related to 't Hooft anomalies.  In
any case, it is hoped that reproducing the answers for the WZ terms
presented here could lead to a better understanding of the 6d \N20\
field theories.

In analogy with ordinary QFT, one might suppose that the coefficient
of the WZ term is some function of only those degrees of freedom which
become massive when $G\rightarrow H\times U(1)$.  E.g. we might try a
function only of $n_W$ which, based on the \sunwz\ case, would then be
the general guess $c(G)-c(H)=3(n_W/2)(n_W/2+1)$.  However, this guess
does not work for the case $G=SO(2N)$ and $H=SU(N)$ in the large $N$
limit: using \cso, we have $c(G)\approx 4N^3$ and $c(H)\approx N^3$,
so we should be getting $c(G)-c(H)\approx 3N^3$; on the other hand,
the guessed formula incorrectly gives $3(n_W/2)(n_W/2-1)\approx
\f{3}{4}N^4$ since $n_W=N(N-1)$.  Based on this failure, it seems that
the coefficient of the WZ term must contain some explicit dependence
on the massless, interacting, $H$ degrees of freedom. If \WZcm\ is
correct, the explicit $H$ dependence is in $\alpha _m$. 

Our conjecture for $c(G)$ for general $G$, based on the $SU(N)$
case and \cso, is 
\eqn\cgisg{c(G)=|G|C_2(G),}
where $C_2(G)$ is the dual Coxeter number, normalized to be $N$ for
$SU(N)$.  This gives $c(SU(N))=N^3-N$, $c(SO(2N))= 2N(N-1)(2N-1)$,
$c(E_6)=(78)(12)=912$, $c(E_7)=(133)(18)=2394$, and
$c(E_8)=(248)(30)=7440$.  A check of \cgisg\ is that it is
a multiple of 6, satisfying
\cquant\ and thus \cdiffq, for all $ADE$ groups $G$.  It also satisfies
the $c$-function condition $c(G)>|G|$ in all cases.

It would be interesting to derive the Hopf-Wess-Zumino term
\WZis, and thus check \cgisg, in the context of IIB string theory on a
$\IC ^2/\Gamma _G$ ALE space, where $\phi ^a$ are the periods of the 3
Kahler forms and two $B$ fields on a blown-up two-cycle.  Since \WZis\
depends only on the angular $\widehat \phi$, the size $|\phi|$ of the
blown-up two cycle can be arbitrarily large. The $C_4\wedge H_3\wedge
H_3^*$ interaction of the 10d IIB string looks promising for leading
to WZ terms.  The $\pi _4(S^4)$ skyrmionic strings should again be
identified with the W-boson strings, which here arise \wcomments\ from
D3 branes wrapped on the blown-up two-cycle.

Decomposing the adjoint of $G$ as $ad(G)\rightarrow W+ad(H)$ for some
representations $W$ (which is the rep of the massive W-bosons, along
with a singlet $=ad(U(1))$) of $G$, the conjectured formula \cgisg\
gives for the coefficient of the WZ term:
\eqn\anguess{c(G)-c(H)=|H|C_2(W)+|W|(C_2(H)+C_2(W)).}
Note that this expression depends explicitly on $H$, via $|H|$ and
$C_2(H)$, and not only on the massive reps in $W$.  This suggests that
an eventual derivation of the WZ term must include effects which
couple the massive degrees of freedom, which are integrated out, to
the massless, interacting, $H$ degrees of freedom.  Assuming \WZcm,
this could be just via $\alpha _m$, the magnetic charge of the
skyrmionic strings in $W_6$ (or membranes in $\Sigma _7$), which would
have explicit $H$ dependence as given by \anguess.  It would be
interesting to directly determine $\alpha _m$ and see if, and how, it
is given as suggested by the above discussion.

\bigskip
\centerline{{\bf Acknowledgments}}

I would like to thank A. Kapustin, A. Manohar, G. Moore, N. Seiberg,
and S. Sethi for discussions.  I would also like to thank E. D'Hoker
and E. Witten for helpful email correspondences.  This work was
supported in part by UCSD grant DOE-FG03-97ER40546 and, while I was a
visitor, by IAS grant NSF PHY-9513835 and Rutgers grant DOE
DE-FG02-96ER40959.  I would like to thank the IAS and Rutgers theory
groups for their support and hospitality.

\listrefs
\end